\documentclass[aps,prd,twocolumn]{revtex4}
\tighten
\input psfig.sty
\usepackage{graphics}

\font\BFd=cmmib10 scaled 1200
\font\BFt=cmmib10 scaled 1200
\font\BFs=cmmib10
 
\def\bb#1{\relax
\ifmmode\mathchoice
{{\hbox{\BFd #1}}}{{\hbox{\BFt #1}}}
{{\hbox{\BFs #1}}}{{\hbox{\BFs #1}}}
\else \mbox{#1} \fi } 

\def\r{{\bb{r}}}
\def\x{{\bb{x}}}
\def\k{{\bb{k}}}
\def\q{{\bb{q}}}
\def\p{{\bb{p}}}
\def\nab{{\bb{$\nabla$}}}

\def\rhocdm{\rho_{\rm cdm}}

\def\go{\mathrel{\raise.3ex\hbox{$>$}\mkern-14mu
\lower0.6ex\hbox{$\sim$}}}
\def\lo{\mathrel{\raise.3ex\hbox{$<$}\mkern-14mu
\lower0.6ex\hbox{$\sim$}}}

\begin{document}

\title{Neutrino Clustering in Cold Dark Matter Halos : Implications for Ultra 
High Energy Cosmic Rays}

\author{Shwetabh Singh}
\email[]{shwetabh@astro.berkeley.edu}
\author{Chung-Pei Ma}
\email[]{cpma@astro.berkeley.edu}
\affiliation{Department of Astronomy, University of California at 
Berkeley,\\ 601 Campbell Hall, Berkeley, CA~94720}

\begin{abstract}

We develop a method based on the collisionless Boltzmann equation to
calculate the gravitational clustering of relic neutrinos in realistic
cosmological models dominated by cold dark matter (CDM) and the
cosmological constant.  This method can be used to estimate the
phase-space distribution of any light particles in CDM halos.  We find
that neutrinos with masses $\agt$ 0.3 eV cluster appreciably in dark
matter halos above the galactic size.  The resulting neutrino
overdensity above the cosmic mean neutrino density increases with both
the neutrino mass and the halo mass, ranging from $\sim 10$ for 0.3 eV
neutrinos in $\sim 10^{13} M_\odot$ halos to $\sim 1500$ for 1.8 eV
neutrinos in $\sim 10^{15} M_\odot$ halos.  We examine the
implications of neutrino clustering for the Z-burst model of ultra
high energy cosmic rays (UHECR), which interprets the observed events
at $E > 4\times 10^{19}$ eV as decay products of Z-bosons from the
resonant scattering between relic and high energy neutrinos and
anti-neutrinos.  We estimate the UHECR energy spectrum for various
neutrino masses towards five of the most massive clusters in the local
universe (within 100 Mpc): Virgo, Perseus-Pisces, Hydra, Centaurus,
and Coma.  The UHECR flux in the Z-burst model is expected to be
significantly higher towards these clusters if $m_\nu\agt 0.3$ eV and
nearly isotropic otherwise.

\end{abstract}

\maketitle

\section{Introduction}

The nature of cosmic rays above the Greisen-Zatsepin-Kuzmin (GZK)
cutoff \cite{GZK} at $\sim$ 4 $\times$ 10$^{19}$ eV is an unsolved
problem in ultra high energy cosmic ray (UHECR) physics \cite{review}.
These events have been reported by the Akemo Giant Air Shower Array
(AGASA) \cite{AG}, Fly's Eye \cite{FE}, Havera Park \cite{HP}, HiRes
\cite{HR}, and Yakutsk \cite{YK} collaborations.  Interactions with
the cosmic background photons $\gamma_{cmb}$ via photoproduction of
pions ($p \gamma_{cmb} \rightarrow p + N \pi, n\gamma_{cmb}
\rightarrow n + N\pi$), photopair production ($p \gamma_{cmb}
\rightarrow p e^{+}e^{-},\gamma \gamma_{cmb} \rightarrow e^{+}e^{-}$),
and inverse Compton scattering ($e^{\pm} \gamma_{cmb} \rightarrow
e^{\pm} \gamma$, $p \gamma_{cmb} \rightarrow p\gamma$) at high
energies \cite{L94} constrain a $\sim 10^{20}$ eV cosmic ray to a few
Mpc for the characteristic lengths of either charged cosmic rays or
neutrons and photons.  More specifically, the attenuation length of
protons above the GZK cutoff is $\sim$ 50 Mpc.  The lack of known
processes to accelerate cosmic rays in small Galactic objects makes
the Galactic origin of these ultra high energy particles unfeasible
\cite{FS95}.  Novel powerful acceleration mechanisms for light nuclei
are required if these energetic particles are produced in nearby
galaxies \cite{A02}.  Exotic particles and dynamics have also been
suggested, but these come with their own difficulties \cite{review}.

One proposed explanation for the UHECRs is the Z-burst model, which
tries to solve the puzzle without invoking new physics beyond the
standard model of particle physics except for neutrino masses.
Several recent experiments \cite{SK601,MAC01,SN01,TR01} have found
evidence for non-zero neutrino mass.  The Z-burst model hinges on the
fact that ultra high energy neutrinos (and anti-neutrinos) produced at
cosmological distances can reach the GZK zone unattenuated.  Their
resonant annihilation on the relic anti-neutrinos (and neutrinos)
produces Z bosons, about 70\% of which decay into hadrons within $\sim
10^{-25}$ sec.  The final state has fifteen pions and 1.35
baryon-antibaryon pairs on average \cite{FMS99}, where the fifteen
pions decay into thirty high energy photons.  The Z boson is highly
boosted ($\sim 10^{10}$) \cite{W99}, resulting in a highly collimated
beam with a half angle of $\sim 10^{-10}$.  This and the fact that the
effect of magnetic fields at such high energies is negligible
\cite{GMT01} ensure a high probability for the protons and photons to
reach the observer if the Z-burst occurs in the direction of the
Earth.  The Z-burst model has been discussed in detail in many papers
\cite{W82,R93,YSL98,GK,CDF01,FKR01,FKR02}.  The resulting cosmic ray
flux has been shown to depend strongly on the density of the relic
neutrinos \cite{YSL98,FMS99,W99,F02,FGSD01}, but the neutrino density
in these calculations has been taken to be either the constant relic
density from the big bang or some ad hoc value.

In this paper we perform a detailed calculation of the neutrino
clustering in the local universe using realistic cosmological models
and apply the results to the Z-burst model for UHECRs.  Since the
current constraints from cosmological observations and laboratory
experiments indicate that the neutrino masses are small ($\lo 2$ eV;
see Sec~II) and the CDM dominates the dark matter density
($\Omega_{\rm cdm} \gg \Omega_\nu$), we do not expect the clustering
of neutrinos to affect significantly that of the CDM.  As a result, it
is not essential to use full scale, time consuming $N$-body
simulations.  Instead, we solve the collisionless Boltzmann equation
for the neutrino phase space distribution in a background potential
given by the universal profile of CDM halos reported in recent high
resolution simulations \cite{NFW96}.  The Boltzmann equation is then
linear in the neutrino density contrast and has tractable integral
solutions.  The advantage of this method over the conventional
$N$-body simulations is that we can obtain the neutrino density
profile much below the resolution scale ($\sim 50$ kpc) of large
cosmological simulations by using as an input the CDM potential
determined from much higher resolution simulations of individual
halos.  Moreover, the computation time required for our approach is
negligible compared with numerical simulations, thereby allowing us to
explore a large parameter space of neutrino masses and dark matter
halo masses.

In Sec~II the relevant Boltzmann equation and the integral solutions
are derived.  In Sec~III results for the clustering of neutrinos for
different neutrino masses and CDM halos are presented and compared
with physical arguments based on neutrino free streaming and the
Tremaine-Gunn constraint \cite{TG79}.  The resulting neutrino density
profiles are also compared with earlier $N$-body simulations
\cite{KKPH96}, which show good agreement.  In Sec~IV the neutrino
overdensity calculation is incorporated in the Z-burst model for
UHECRs, where we make realistic predictions for the UHECR energy
spectrum for different neutrino masses.  We estimate the level of
anisotropy in the UHECR flux by examining lines of sight towards five
of the most massive clusters (Virgo, Perseus-Pisces, Hydra, Centaurus,
and Coma) in the local universe (within 100 Mpc) where neutrinos are
likely to be most clustered.

\section{Boltzmann Equation for Neutrino Clustering in CDM Halos}

In this section we develop an approach based on the collisionless
Boltzmann equation to study how massive neutrinos cluster
gravitationally in realistic cosmological models.  We start by noting
that the median velocity of unclustered background neutrinos of mass
$m_\nu$ (in eV) at redshift $z$ is
\begin{equation}
   \bar v = 161 (1+z) m^{-1}_\nu \,  {\rm km \, s^{-1}}\,.
\label{medianV}
\end{equation}
This implies that light neutrinos ($m_{\nu}\alt 2$ eV) do not accrete
significantly onto CDM protoclusters until $z\sim 3$ because the
neutrino thermal velocities are greater than the velocity dispersion
of a typical cluster or supercluster.  We are then faced with the more
tractable problem of how neutrinos cluster in the potential well of an
existing CDM halo.  Our approach is to use the collisionless Boltzmann
equation for the neutrino phase space density $f$ and follow its
evolution in a background CDM potential given by the approximate
universal profile obtained in high resolution simulations of
individual halos \cite{NFW96}.  Note that the CDM potential is
time-dependent in general.  Earlier work has used the Boltzmann
approach to study how neutrinos cluster around point masses in the
context of cosmic string seeded galaxy formation \cite{BKT87,BW88}.
This method allows us to calculate the neutrino density profiles in
the inner part of the cluster ($\lo 10$ kpc) that can not be resolved
by large cosmological simulations.  This will be seen to be important
in the Z-burst model where a significant contribution to the cosmic
ray flux comes from the inner regions of the halo.

In the Newtonian approximation and in physical coordinates, the
collisionless Boltzmann equation takes the form
\begin{equation}
    \frac{\partial f}{\partial t} +\dot{\r} \cdot \nab_r f 
    +\dot{\p} \cdot \nab_p f=0. 
\end{equation}
Rewriting it in conformal time $d\tau=dt/a$ and in comoving position
$\x=\r/a$ and momentum $\q=a\p-m_\nu \dot{a}\r$, we obtain
\begin{eqnarray}
   & &\frac{1}{a} \frac{\partial f}{\partial \tau}  
     + \frac{\q}{m_\nu a^2}\cdot \nab_x f - m_\nu \ddot{a}a \x \cdot \nab_q f  
	\nonumber \\ 
    & & - m_\nu \nab_x \Phi \cdot \nab_q f=0 \,,
\label{boltz0}
\end{eqnarray}
where the Newtonian gravitational potential $\Phi$ obeys
$\dot{\p}=-m_\nu\nab \Phi$.  At the time of decoupling the neutrino
phase space density is given by the thermal Fermi-Dirac distribution
$f_{0}(q)\propto (e^{q/T_{\nu,0}} +1)^{-1}$ where
$T_{\nu,0}=(4/11)^{1/3} T_{\gamma,0}=1.676 \times 10^{-4}$ eV is the
temperature of the neutrino background today.  Gravitational
clustering distorts the spatially uniform $f_0$, so we write the full
distribution as
\begin{equation}
	f(\x, \q, \tau)= f_{0}(q) + f_{1}(\x,\q,\tau)\,.
\end{equation}
The gravitational potential can also be written as 
\begin{equation}
	\Phi(\x,\tau)=\Phi_{0}(\x,\tau) + \Phi_{1}(\x,\tau) \,,
\end{equation}
where $\Phi_0$ is related to the mean background comoving density
$\bar\rho=\bar\rho_{\rm cdm} + \bar\rho_\nu$ by $\nab_{x} \Phi_{0}=
\frac{4 \pi}{3} G\bar\rho a^2 \x$, and $\Phi_{1}$ is determined by the
density contrast of both CDM and neutrinos in the halo:
\begin{eqnarray}
     \nabla^2_x \Phi_1 &=& 4 \pi G a^2 (\delta\rhocdm  + \delta \rho_\nu)
\end{eqnarray}
Eq.~(\ref{boltz0}) then becomes
\begin{eqnarray}
   & &\frac{1}{a} \frac{\partial f_{1}}{\partial \tau} 
	+ \frac{\q}{m_\nu a^2} \cdot \nab_{x} f_{1} 
	- m_\nu \nab_{x} \Phi_{1} \cdot\nab_q f_0 \nonumber \\
	& &- m_\nu \nab_{x} \Phi_{1} \cdot \nab_{q} f_{1}=0\,, 
\label{boltz}
\end{eqnarray}	
where we have used the Friedmann equation $\ddot{a}=-\frac{4 \pi}{3} G
\bar{\rho} a$, which gives $\ddot{a} a \x + \nabla_x \Phi_0 =0$.  We
note that eq.~(\ref{boltz}) is the full Boltzmann equation and no
approximation has been made thus far.  It is generally a nonlinear
equation in $f_1$ where $\Phi_1$ is related to $f_1$.  For our
problem, however, $\Phi_1$ is mostly determined by the CDM whose
potential has a well known, pre-determined form.  Eq.~(\ref{boltz}) is
then linear in $f_1$ and much easier to solve.

Furthermore, eq.~(\ref{boltz}) has a simple integral solution if the
fourth term is neglected.  For example, in earlier calculations that
examined neutrino clustering onto point masses seeded by cosmic
strings, this term was dropped in order to simplify the calculation
\cite{BKT87,BW88}.  We will also drop this term, but we justify this
approach in two ways.  First, we note that dropping this term requires
$\nab_q f_1 < \nab_q f_0$ and not $f_1 < f_0$. The former is generally
a less restrictive condition and can be satisfied even if $f_1$ is
much larger than $f_0$ because on dimensional grounds, we have
\begin{equation}
   {\nab_q f_1 \over \nab_q f_0} \sim {f_1/\sigma_v \over f_0/ \bar v}
   \sim { \delta\rho_\nu / \sigma_v^4 \over \bar\rho_\nu / \bar v^4 }
          \sim \delta_\nu \left( { \bar v \over \sigma_v} \right)^4,
 \label{dim}         
\end{equation}
where $\sigma_v$ is the velocity dispersion of the gravitational
potential $\Phi_1$ and $\bar v$ is the median neutrino thermal
velocity in eq.~(\ref{medianV}).  Since only neutrinos with $\bar v
\ll \sigma_v$ are cold enough to fall into the gravitational wells, we
expect the ratio $\nab_q f_1/ \nab_q f_0$ to be much smaller than
$f_1/f_0$, thereby making it easier to justify the dropping of the
fourth term in eq.~(\ref{boltz}).  (For example, ${\nab_q f_1 / \nab_q
f_0} \lo 0.2$ for $\sim 1$ eV neutrinos in $\sim 10^{14} M_{\odot}$
halos.)  Eq.~(\ref{dim}) further indicates that the neutrino
overdensity is larger than $f_1/f_0$ by $(\sigma_{v} / \bar{v})^{3}$,
a large factor in highly clustered regions.  This explains
qualitatively the large overdensity found in our numerical results to
be presented in Sec III.  In the next section we also provide further
justification for ignoring the $\nab_q f_1$ term by comparing our
results with the neutrino density profiles of two halos obtained from
earlier numerical simulations.

Following \cite{BKT87}, we convert eq.~(\ref{boltz}) into an ordinary
differential equation by going into Fourier space and using a new time
variable $d\eta=d\tau/a$:
\begin{equation}
   \frac{\partial \tilde{f_{1}}}{\partial \eta}  
  + \frac{i \k \cdot \q}{ m_\nu} \tilde{f_{1}} 
  + {i m_\nu \over k^2} 4 \pi G a^4 \tilde\rho \,\k\cdot \nabla_q f_0 = 0\,,
\label{ec10}
\end{equation}
where $\tilde{f_1}$ and $\tilde\rho$ are the Fourier transforms
of $f_1$ and $\rho$.  The solution is 
\begin{eqnarray}
   \tilde f_1 (\k, \q,\eta) & = & - {i m_\nu\over k^2} 4 \pi G
    \int^{\eta}_{\eta_{0}} 
	d \eta' e^{-i \k \cdot \q (\eta-\eta')/m_\nu}\nonumber \\   
   & & \times a^4(\eta') 
        \tilde{\rho}(\k,\eta')\, \k \cdot \nab_{q} f_0 \,,
\label{f1}
\end{eqnarray}
where we have taken the initial neutrino phase space to be
homogeneous, i.e., $f_1(\eta_0)=0$ and $f(\eta_0)=f_0$.

The comoving neutrino number density is given by
\begin{equation}
   \tilde{n}_{\nu}(\k,\eta)-\bar{n}_{\nu}=\frac{2}{h_p^3} 
	\int d^3 \q \tilde{f_{1}}(\k, \q, \eta) \,,
\end{equation}
which can be obtained from eq.~(\ref{f1}) after integration by
parts in $q$ and integration over angles:
\begin{eqnarray}
   \tilde{n}_{\nu}(\k,\eta)-\bar{n}_{\nu} & = & \frac{32\pi^2 G m_{\nu}}
	 {h_p^3 k}\int_{\eta_{0}}^{\eta} d\eta' 
     a^{4}(\eta')   \tilde\rho(k,\eta')\nonumber \\ 
   & & \times \int_{0}^{\infty} dq\, q \, 
	{\sin [k\, q (\eta-\eta')/m_\nu] \over e^{q/T_{\nu,0}}+1 } \,,
\label{delnu}
\end{eqnarray} 
where $\bar n_\nu \approx 112$ cm$^{-3}$ is the cosmic mean comoving
number density of one species of light neutrinos and antineutrinos,
and $h_p$ is the Planck constant.  Eq.~(\ref{delnu}) is the main
equation that we will solve in this paper.  It is a Volterra integral
equation of the second kind that has the form $f(t)=\int^t_a ds\,
K(t,s) f(s) + g(t)$ and can be solved with the trapezoidal rule.  It
describes how neutrinos of a given mass $m_\nu$ cluster around a
realistic CDM halo as a function of time.  The density of the halo at
a given time, $\tilde{\rho}$, should generally be the sum of the CDM
and the neutrino components, but as we verify below, using the CDM
potential alone is a very good approximation for the cosmological
models of current interest.

\section{Results for Neutrino Clustering in CDM Halos}

In this section we present results for $n_{\nu}$ computed from
eq.~(\ref{delnu}).  We choose to integrate eq.~(\ref{delnu}) from
$z=3$ to 0; the results differ by only about 5\% if the initial
redshift is pushed back to 5 because the neutrinos do not cluster
appreciably at such early times as discussed above.  We also need to
specify the cosmological models and neutrino masses.  For the
cosmological parameters, we use the currently favored critically-flat
model with matter density $\Omega_m=\Omega_{\rm cdm} +
\Omega_\nu=0.35$, cosmological constant $\Omega_\Lambda=0.65$, and
Hubble parameter $h=0.7$.  Variations in these parameters at 10 to
20\% level are not expected to alter our neutrino results
significantly since the effect on the halo potential $\Phi_1$ is
small.  For the neutrino masses, we consider four different models in
which three models assume three degenerate massive species with masses
0.6, 0.3, and 0.15 eV respectively, and one model with one massive
species with mass 1.8 eV.  The corresponding density parameter in
neutrinos is $\Omega_{\nu}\approx 0.04, 0.02, 0.01$ and 0.04,
respectively, all much smaller than $\Omega_{\rm cdm}$.  This range of
neutrino masses is chosen to span the current cosmological and
laboratory limits.  The most recent cosmological constraint comes from
the galaxy clustering power spectrum of 2-Degree-Field Redshift
Survey, which places an upper limit of 1.8 eV on the sum of the
neutrino masses \cite{E02}.  The Super-Kamiokande experiment
\cite{SK601} provides strong evidence for oscillations between
neutrino species with a mass difference of $\delta m^{2}= (1-8) \times
10^{-3}$ eV, giving a minimum mass of $\approx 0.07$ eV if the
neutrino masses are hierarchical. The choice of three degenerate
neutrino masses is based on indications that if any of the mass
eigenvalues is above 0.1 eV then all three masses are above 0.1 eV and
almost degenerate \cite{BB02}.

Although the final neutrino density profile will depend strongly on
the CDM profile, we do not expect the reverse to hold: CDM density
will not be much affected by the clustering of neutrinos because all
the models considered in this paper have small
$\Omega_{\nu}$/$\Omega_{\rm cdm}$, so the CDM dominates the
gravitational potential of a dark matter halo.  We are therefore
justified in using the universal CDM profile determined from the high
resolution pure CDM simulations \cite{NFW96} as the input:
\begin{equation}
   \rhocdm (r)= \frac{\bar\rho\, \bar\delta\, r_s^3}{r(r+r_s)^2}\,,
\label{nfw}
\end{equation}
where $\bar{\delta}$ and $r_{s}$ are given in terms of the
concentration parameter \cite{Bet01,MF00}
\begin{eqnarray}
    c &=& \frac{9}{1+z} \left(\frac{M}{1.5 \times 10^{13} 
	h^{-1} M_{\odot}} \right)^{-0.13}\,, \nonumber\\
   \bar\delta &=&  \frac{200 c^{3}}{3 \left[\ln(1+c)- c/(1+c) \right]}\,,
	\label{para}\\
   r_{s} &=& \frac{1.63 \times 10^{-5}}{\Omega_{m}^{1/3} c} 
 \left(\frac{M}{h^{-1} M_{\odot}} \right)^{1/3} h^{-1} {\rm Mpc} 
   \nonumber\,.
\end{eqnarray}
We evaluate the mass density $\rho$ on the right hand side of
eq.~(\ref{delnu}) exactly by adding the CDM density given above and
the neutrino density computed from previous time steps.  We did find
that approximating $\rho$ with the CDM profile alone (i.e. ignoring
the neutrino contribution to the total potential) changes our results
by no more than 10\% for the light neutrino masses and cosmological
parameters considered in this paper.  We also tested the simplifying
assumption made in \cite{BKT87}, which allowed them to reduce the
integrals in eq.~(\ref{delnu}) analytically to a single integral by
using $(e^{q/T_{\nu,0}}+1)^{-1} = A\,e^{-q/T_{\nu,0}}$, i.e., by
assuming a Maxwell-Boltzmann instead of Fermi-Dirac distribution.  We
found this simplification to cause a $\sim$ 20 \% difference at small
scales, so we do not use this approximation here.

Before presenting our results for the realistic models above, we first
conduct a comparison study by checking our results for the neutrino
density profile against those in the numerical simulations of Ref.
\cite{KKPH96}, which investigated the clustering of CDM and neutrinos
in two flat cosmological models: $\Omega_{\rm cdm}=0.8$ and
$\Omega_{\nu}=0.2$ with two species of 2.3 eV neutrinos, and
$\Omega_{\rm cdm}=0.7$ and $\Omega_{\nu}=0.3$ with one species of 7 eV
neutrinos.  Both models assumed $h=0.5$.  These models are no longer
consistent with current observations, but the simulation results
provide a useful tool for us to test the validity of our Boltzmann
approach.  For a fair comparison, we use the CDM halo profile found in
\cite{KKPH96} as an input:
\begin{equation}
   \rhocdm (r)= \frac{C}{r(r+R)^{\alpha}} \,,
\label{kofman}
\end{equation}
where $R$ and $\alpha$ are 0.3 Mpc and 1.5 for the $\Omega_\nu=0.2$
model and 0.11 Mpc and 1.1 for the $\Omega_\nu=0.3$ model.  We note
that their outer profile is shallower than eq.~(\ref{nfw}) from the
pure CDM simulations \cite{NFW96}.  In Ref.~\cite{KKPH96} this
difference was attributed to the differing spectral index $n$ for the
matter power spectrum of the neutrino models at the mass scale of the
simulated halos ($\approx 1.3 \times 10^{15}$ M$_{\odot}$): $n\simeq
-1.36$ for $\Omega_{\nu}$=0.2 and $n\simeq -1.53$ for
$\Omega_{\nu}$=0.3.  This argument, however, appears inconsistent with
the near universal nature of the halo density profile reported in
Ref.~\cite{NFW96}.  A better understanding for the origin of halo
profiles should help resolve this issue.

Fig.~1 compares the ratio of the neutrino and CDM density profiles
from our approach vs. the two simulated halos in Ref.  \cite{KKPH96}.
For both halos we have used the same cosmological parameters, CDM
profiles, and halo mass in our Boltzmann approach as in their
simulations.  We find a good agreement between the two methods for the
outer parts of the cluster; whereas our results are lower by about
50\% in the inner parts.  This discrepancy may be due to our
neglecting the fourth term in eq.~(\ref{boltz}), or due to statistical
fluctuations in the substructure in their simulations.  (Only two
simulated halos are presented in Ref. \cite{KKPH96}.)  The Boltzmann
approach used here also allows us to explore how neutrinos respond to
different CDM potentials.  As an illustration of this, we show in
Fig.~1 how the neutrino profile changes when the input CDM profile is
changed from eq.~(\ref{kofman}) to the higher resolution profile of
eq.~(\ref{nfw}).  We conclude from Fig.~1 that we can obtain a
reasonable estimate for neutrino clustering using eq.~(\ref{delnu})
instead of full scale $N$-body simulations.

We now turn to our results for the realistic cosmological models and
neutrino masses given at the beginning of this section.  The four
panels in Fig.~2 show the neutrino overdensity $n_\nu/\bar n_\nu$
computed from eq.~(\ref{delnu}) for four models of neutrino masses.
The more massive neutrinos clearly cluster more because of their lower
thermal velocities.  Within each panel, the four curves illustrate how
$n_\nu$ increases with halo masses from $10^{12}$ to $10^{15} M_\odot$
as a result of the deeper halo gravitational potentials.  The growth
of $n_\nu$ in the inner parts of the halo, where it is almost
independent of $r$, is illustrated in Fig.~3 for 0.7 and 0.4 eV
neutrinos in $10^{15}, 10^{14}$, and $10^{13} M_\odot$ halos.  Most of
the clustering is seen to occur at low redshifts.

Unlike the CDM density that continues to rise towards the inner part
of a halo as $\rho\propto 1/r$, all curves in Fig.~2 show that
$n_{\nu}$ flattens out at some radius.  Similar features were also
seen in Ref.~\cite{BW88} for neutrinos clustered around cosmic
strings.  This relative suppression in the neutrino vs. CDM
overdensity reflects neutrino free streaming, which dampens and
retards perturbation growth on small length scales due to phase
mixing.  The neutrino damping scale, $R_d$, can be characterized by
the length scale above which neutrinos behave like the CDM.  The
standard method to solve the Boltzmann equation for a fluid with
pressure invloves the transformation of the Boltzmann equation into an
infinite hierarchy of velocity moment equations \cite{MB95}, where the
lowest three moments with $l=0,1$, and 2 correspond to the density,
velocity, and shear of the fluid.  The choice of the truncation of the
hierarchy depends on the physical properties of the fluid and the
length scales.  For CDM, for example, all modes above $l=1$ are zero.
The parameter $R_d$ gives the scale above which the Boltzmann
hierarchy for neutrinos can be truncated at $l=1$ (as for the CDM),
and below which more $l$-modes must be included to compute the
neutrino damping effect accurately.  This scale is given by
\cite{DGS96}
\begin{equation}
  R_d(\tau) \equiv {\tau \over\sqrt{1+ [ a(\tau)/ a_{nr} ]^2} } \,,
\label{free}
\end{equation}
where $a_{nr}\simeq 3T_{\nu,0} /m_{\nu}$ is the expansion factor at
which the neutrinos become non-relativistic.  For the cosmological
models considered in this paper, the scale is $R_d(z=0)\approx
5.8/m_\nu$(eV) Mpc.  From Fig.~1, we indeed find the ratio
$\rho_{\nu}/\rho_{cdm}$ to be about the cosmic mean value
$\Omega_\nu/\Omega_{cdm}$ at scales above $R_d$ and to decrease
gradually at smaller radii (top panels), with a final flattening in
the neutrino overdensity at $r\sim 0.1\,R_d$ (bottom panels).  Fig.~2
shows that the radius at which $\delta_\nu$ flattens out depends
weakly on the mass of the CDM halo.  It occurs at smaller radii for
less massive CDM halos primarily because the lower mass halos provide
shallower potential wells.  The damping scale $R_d$ of
eq.~(\ref{free}) is to be contrasted with the neutrino free streaming
distance, which is typically defined as the comoving distance
traversed from the time of neutrino decoupling to $a_{nr}$:
$\lambda_{fs} \equiv \int_{\tau_{i}}^{\tau_{nr}} {d\tau'}/{a(\tau')}
\simeq 600/m_{\nu}$(eV) Mpc \cite{KT90}.  The distance $\lambda_{fs}$
reflects the global streaming motion of neutrinos but not the local
clustering properties of neutrinos around CDM after they become
non-relativistic.

Another way to understand the results in Figures~1 and 2 is to compare
the thermal velocities of neutrinos with the velocity dispersions of
the CDM halos: neutrinos can cluster significantly only if their mean
thermal velocity in eq.~(\ref{medianV}) is below the typical velocity
of the host CDM halo.  Fig.~4 compares these two characteristic
velocities for a range of neutrino masses and halo masses.  Since the
NFW profile specifies only the spatial and not the velocity
distribution of the CDM halo particles, two velocity ellipsoids are
shown for comparison: isotropic, which is appropriate near the center
of the halo, and the more radial distribution
$\beta=1-v_t^2/v_r^2=0.5$, which is appropriate for the outer regions.
Fig.~4 illustrates that $< 0.15$ eV neutrinos are too hot to be
captured significantly by $\alt 10^{14} M_\odot$ halos, while the more
massive neutrinos can fall into progressively lower mass halos, a
result consistent with that shown in Fig.~2.

Our results for neutrino clustering can be compared with the
Tremaine-Gunn bound \cite{TG79}, which gives an upper limit on the
neutrino density in the core of a halo based on the argument that the
maximum coarse grained phase space density can not exceed the maximum
initial phase space density due to phase mixing.  Their results are
not directly applicable to our problem because in their derivation,
neutrinos are assumed as the sole constituent of dark matter, and the
coarse grained neutrino distribution is assumed to be
Maxwell-Boltzmann instead of Fermi-Dirac for computational
convenience.  More recent work \cite{KTB96} has extended the
derivation to models including both CDM and neutrinos and obtained
$\rho_{\nu} \le |2\Phi|^{3/2}m_{\nu}^{4}/12 \pi^4$, where $\Phi$ is
the gravitational potential of the system.  For the NFW profile, we
find
\begin{eqnarray}
  \frac{n_{\nu}}{\bar{n}_{\nu}}(r;m_{\nu},r_{s},\bar{\delta}) & \alt & 40 
  \left(\frac{m_{\nu}}{{\rm eV}}\right)^{3}
  \left(\frac{r_{s}}{{\rm Mpc}}\right)^{3} \bar{\delta}^{3/2} 
	\nonumber\\ & & 
 \times \left[\frac{r_{s}}{r}
  \ln\left(\frac{r}{r_{s}}+1\right)\right]^{3/2} \,,
\label{TG}
\end{eqnarray}
where $r_{s}$ and $\bar{\delta}$ are the CDM halo parameters given by
eq.~(\ref{para}).  For 1.8 eV neutrinos, for example, this formula
gives $n_{\nu}/\bar{n}_{\nu} < 3.9 \times 10^4, 3.2\times 10^5,
2.7\times 10^6,$ and $2.3 \times 10^7$ for $10^{12}, 10^{13}, 10^{14}$
and $10^{15} M_\odot$ halos, respectively, at the scale radius $r_s$.
One can see that this constraint is satisfied by at least three orders
of magnitude for all neutrino overdensities in Figures~1 and 2.

\section{Implications for Ultra High Energy Cosmic Rays}

In this section we apply the neutrino clustering results from Sec~III
to the Z-burst model for UHECRs.  No previous work on the Z-burst
model has included realistic calculations for $n_{\nu}$.  Instead, the
value of $n_{\nu}$ has been chosen based on certain observational
constraints \cite{FKR02} or physical arguments \cite{W99,YSL98} and
has differed greatly from $n_{\nu} \sim (1$ - $10^5) \bar{n}_{\nu}$.
For instance, in \cite{FKR02}, it is inferred from the CDM
distribution in our local universe, but the large smoothing scale
$\sim$ 20 Mpc assumed in the calculation results in $n_{\nu} \sim
\bar{n}_{\nu}$.  In contrast, our results from Sec~III show that
$n_{\nu}$ can be $\gg \bar{n}_{\nu}$ in the inner $\sim 1$ Mpc of CDM
halos.  In Refs. \cite{W99,YSL98}, $n_{\nu}$ is approximated based on
phase-space arguments similar to that of \cite{TG79}.  While this
approach gives an upper bound on the neutrino clustering, the actual
overdensity can be significantly less, as we have discussed in the
previous paragraph.  In addition, the neutrino clustering scale of
$\approx$ 5 Mpc assumed in \cite{YSL98} is much larger than what we
find in our calculations.  Our method gives specific predictions for
$n_{\nu}$ as a function of halo radius, halo mass, and neutrino mass.

To estimate the cosmic ray flux we follow the standard assumption in
the Z-burst model that the UHECRs above the GZK cutoff are produced by
the resonant $\nu \bar\nu$ scattering, while the lower energy events
are explained by protons originating from a uniform distribution of
extragalactic sources.  The latter appears consistent with the
isotropic distribution of $E< 4\times 10^{19}$ eV events detected in
AGASA and HiRes \cite{BM99}.  We compute the cosmic ray flux (in (eV
m$^2$ s sr)$^{-1}$) from the Z-burst model with \cite{FKR02}
\begin{widetext}
\begin{eqnarray}
  F_{Z}(E) & = &  \int_{0}^{\infty} {\rm d}E_{p} \int_{0}^{R_{max}}
   {\rm d}r \left[\int_{0}^{\infty} {\rm d}E_{\nu_{i}}
   F_{\nu_{i}}(E_{\nu_{i}},r)n_{\bar{\nu}_{i}}(r)+\int_{0}^{\infty}
   {\rm d}E_{\bar{\nu}_{i}}
   F_{\bar{\nu}_{i}}(E_{\bar{\nu}_{i}},r)n_{\nu_{i}}(r) \right]
   \nonumber\\ & & \times \sigma_{\nu \bar{\nu} }(s) \, {\rm Br}(Z
   \rightarrow {\rm hadrons}) \frac{{\rm d} N_{p+n}}{{\rm d} E_{p}}
    \left| \frac{\partial P_{p}(r,E_{p};E)}{\partial E} \right| \,.
\label{fluxZ1}\
\end{eqnarray}
\end{widetext}
Here $F_{\nu_{i}}(E_{\nu_{i}},r)$ is the flux of ultra high energy
neutrinos with energy $E_{\nu_{i}}$ at distance $r$ and
$n_{\nu_{i}}(r)$ is the physical number density of the relic
neutrinos.  (The repeated index $i$ is summed over different neutrino
species.)  The particle interactions are described by the cross
section $\sigma_{\nu \bar{\nu}}(s)$ for the Z-boson production at the
center-of-mass energy $s=2 m_\nu E_\nu$, and by the branching ratio
${\rm Br}(Z \rightarrow {\rm hadrons})=69.89 \pm 0.07 $ \% for the
subsequent cascade of the Z-boson into hadrons \cite{PDGC00}.  The
factor ${\rm d} N_{p+n}/{\rm d} E_{p}$ gives the energy distribution
of the produced protons and neutrons.  The subsequent proton
propagation is specified by $P_{p}(r,E_{p};E)$, which gives the
probability that a proton created at distance $r$ with energy $E_{p}$
arrives at Earth with an energy greater than $E$.  It measures the
amount of proton energy degradation due to the resonant
photoproduction of pions and other processes discussed in Sec~I.
Specific values of $P_p$ has been calculated in \cite{FK01} for the
range of parameters considered in this paper.  We do not include in
our UHECR flux estimate the contributions from the photons produced in
the Z-decay because experimental data suggest that less than $50 \%$
of the cosmic rays above $4 \times 10^{19}$ eV are photons at the $95
\%$ confidence level \cite{AHVWZ02,AG2}.  Typical physical mechanisms
used to explain the suppressed photon contributions are large
universal radio background and sufficiently strong extragalactic
magnetic fields ($\agt 10^{-9}$ Gauss) \cite{FKR02, PB96}.  The study
of the effects of these parameters on the UHECR flux due to the
Z-burst model can be a subject of future work.

A key feature of the Z-burst model is that the cross section
$\sigma_{\nu \bar{\nu}}(s)$ for $\nu\, \bar\nu \rightarrow Z^0$ is
enhanced by several orders of magnitude near the resonant energy in
the rest frame of the relic neutrinos \cite{W82}
\begin{equation}
   E^{res}_{\nu_{i}}= { M_Z^2 \over 2 m_{\nu_i}} = 4.2 \times 10^{21} 
	{\rm eV} \left(\frac{1\, {\rm eV}}{m_{\nu_{i}}}\right) \,,
\label{energy}
\end{equation} 
where $M_Z$ is the mass of the Z boson.  The flux in
eq.~(\ref{fluxZ1}) to a good approximation therefore depends only on
the neutrino resonant energy and not on the slope of the incident high
energy neutrino spectrum.  Eq.~(\ref{fluxZ1}) can then be written as
\cite{FKR02}
\begin{eqnarray}
   F_Z (E) & = &\bar{\sigma}_{\nu \bar{\nu}}
   F_{\nu_{i}}(E_{\nu_{i}}^{res}) \int_{0}^{\infty} dE_{p}
   \int_{0}^{R_{max}} {dr}\nonumber \\ 
   & & \times n_{\nu_{i}}(r) \,
	\, Q_{p}\left(y=\frac{4 m_{\nu} E_{p}}{M_{Z}^{2}}\right)
	\nonumber\\
    & & \times \left| \frac{\partial P_{p}(r,E_{p};E)}{\partial E}
	\right| \,,
\label{fluxZ2}\
\end{eqnarray}
where $n_{\nu_i}(r)$ is the physical number density of neutrinos and
antineutrinos at the Z-burst site at radial distance $r$,
$\bar{\sigma}_{\nu\bar{\nu}}=40.4$ nb is the cross section for $\nu\,
\bar\nu \rightarrow Z^0$ averaged over the width of the resonance, and
$F_{\nu_{i}}(E_{\nu_{i}}^{res})$ is the incident flux of ultra high
energy neutrinos at the resonant energy.  The function $Q_{p}$ is the
boosted momentum distribution from hadronic Z decays and can been
calculated from experimental data \cite{FKR02}.  It has a fairly broad
peak at $y \approx 10^{-2}$ and falls off approximately as $y^{-7}$
for $y \agt 0.5$.  The neutrino flux $F_{\nu_{i}}(E_{\nu_{i}}^{res})$
remains a free parameter in the Z-burst calculation since no
successful astrophysical model yet exists to explain the production of
$\agt$ 10$^{21}$ eV neutrinos \cite{VMO97,WB,GV02}.  We do not attempt
to model the effect of source evolution in our calculations since it
is again an unknown quantity and is easy to incorporate once its
nature is known.

We first present results for the cosmic ray spectrum $F(E)$ in the
Z-burst model ignoring the spatial clustering in the neutrinos, i.e.,
we assume $n_{\nu}=\bar{n}_{\nu}$ in eq.~(\ref{fluxZ2}).  This
assumption underestimates the flux in the Z-burst model, but we
include the results here for comparison since this is a common
assumption made in several Z-burst calculations \cite{FKR02,W99,GV02}.
Fig.~5 shows the sensitivity of $E^3\,F(E)$ on the neutrino masses.
The flux is higher at high energies for smaller $m_\nu$ because the
momentum distribution $Q_{p}$ of the decay particles peaks at a higher
energy for smaller $m_\nu$.  For a given $m_\nu$, $E^3\,F(E)$
decreases rapidly at $E\go 10^{21}$ eV because $Q_{p}$ falls off as
$\sim y^{-7}$ for $y \agt 0.5$.  The integration is carried out to a
maximum distance of $R_{max}=2000$ Mpc, but our results are
insensitive to this choice as long as $R_{max}$ is sufficiently beyond
the GZK zone of $\sim 50$~Mpc.

For ease of comparison, the curves in Fig.~5 are all normalized to the
same incident neutrino flux of $F_{\nu_{i}}(E_{\nu_{i}}^{res})=1.7
\times 10^{-35}$ (eV m$^2$ s sr)$^{-1}$ for each of the three neutrino
flavors.  (For the one flavor $0.07$ eV model, the assumed flux is 3
times higher.)  We do not attempt to determine this value by
performing statistical fits to data because the UHECR spectrum from
AGASA (square symbols) and HiRes (triangle symbols) disagree in both
amplitude and shape.  We do note that for models that have three
degenerate neutrino masses of $m_{\nu_i}\lo 1$ eV, this value for the
neutrino flux is consistent with the existing upper bound from the
Goldstone Lunar Ultra-high energy neutrino Experiment (GLUE)
\cite{GLUE}.  The 0.07 eV model shown in Fig.~5, however, would need
to be lowered by a factor of $\sim$ 4 in order to satisfy the GLUE
upper limit.  A better understanding of systematic effects in the GLUE
experiment is needed before their results can be used to rule out
models.

For comparison, the dotted curve in Fig.~5 shows the cosmic ray flux
for protons originating from a uniform distribution of extragalactic
sources with a constant comoving density.  It is computed from
\begin{eqnarray}
   F_{EG}(E) & = &  \int_{0}^{\infty} dE_{p} 
  \int_{0}^{R_{max}} \frac{dr}{R_{max}} [1+z(r)]^3 \nonumber \\ 
  & & \times F_p (E_p) \left| \frac{\partial P_{p}(r,E_{p};E)}{\partial E}
	\right| \,,
\label{fluxEG}
\end{eqnarray}
where the unknown proton injection energy spectrum $F_p(E_p)$ is
typically assumed to be a power law: $F_p(E_p)=\epsilon^{-1} A
E_p^{-\beta}$.  The shape of $F_{EG}$ depends on the injection
spectrum $F_p$, but for definiteness, we have assumed $\beta=2.4$ and
$A=5.98 \times 10^{31}$ (and an upper energy cutoff of $E_p=10^{23}$),
which are found to be the best fit values \cite{FKR02} to the existing
cosmic ray data that have a total experimental exposure of $\epsilon
\approx 8 \times 10^{16}$ m$^{2}$ s sr.  The GZK cutoff is clearly
seen at $\sim$ 4 $\times$ 10$^{19}$ eV in the dotted curve.  The flux
rises beyond $\sim 4 \times 10^{20}$ eV because the photoproduction of
pions is a resonant process where the cross section peaks at $E_{p}
\sim 2.3 \times 10^{20}$ eV \cite{GZK} and decreases at higher
energies, allowing a larger fraction of protons to reach us.

The predictions for the UHECR spectrum change significantly when we
incorporate the neutrino overdensity computed in Sec~III.  To make
realistic estimates for our local universe, we consider five lines of
sight towards five of the most massive nearby clusters: Virgo,
Centaurus, Hydra, Perseus-Pisces, and Coma, where the highest
overdensity of neutrinos are expected.  The distance, mass, rough
angular extent, and equatorial coordinates of each of the clusters are
listed in Table 1 \cite{VIZ,NED}.  The cluster masses are taken from
{\it http://cfa-www.harvard.edu/$\ \tilde{}$huchra/clusters}, where
they are estimated from galaxy velocities and the virial theorem.  We
caution that these values have large error bars.  The mass of the
nearest cluster Virgo \cite{Bet94,TS,SBB99,FSSB01}, for example, has
been estimated to be $1.5 - 6 \times 10^{14}$ M$_{\odot}$ based on
X-ray emission measurements by ROSAT \cite{Bet94}, to $1.5 \times
10^{15}$ M$_{\odot}$ based on the relativistic Tolman-Bondi method
\cite{FSSB01}.  (The Tolman-Bondi model is based on analytic solutions
to the Einstein field equations for spherically symmetric pressure
free overdensities in a homogeneous universe \cite{FSSB01,ELTPF99}.)

\begin{table*}
\caption{Parameters of five nearby clusters}
\begin{ruledtabular}
\begin{tabular}{cccccc}
Name & Distance(Mpc) & Mass(M$_{\odot}$) & Angular Radius($^{\circ})$ & RA(h 
min) & Dec($^{\circ}$ min)\\
\hline
Virgo & 15 & 7.9 $\times$ 10$^{14}$ & 5 & 12 29.6 & +11 49\\
Centaurus & 43 & 1.3 $\times$ 10$^{15}$ & 1.5 & 12 46.1 & $-$41 02\\
Hydra & 53 & $4.6 \times 10^{14}$ & 2 & 10 34.5 & $-$27 16\\ 
Perseus-Pisces & 76 & 5.5 $\times$ 10$^{15}$ & 7 & 03 15.3 & +41 20\\
Coma & 99 & 1.7 $\times$ 10$^{15}$ & 2.5 & 12 57.4 & +28 15\\
\end{tabular}
\end{ruledtabular}
\end{table*}

Fig.~6 shows our predictions for the cosmic ray flux (in (eV m$^2$ s
sr)$^{-1}$) towards these five lines of sight for four different
neutrino masses.  Along each line of sight, high energy neutrinos from
extragalactic sources are assumed to traverse a uniform sea of
background neutrinos plus an overdensity of background neutrinos
centered at the location of the given cluster, where $n_\nu$ is
computed from eq.~(\ref{delnu}) for the mass of the cluster.  We also
include in our calculation a local $n_\nu$ for the Local Group of mass
$4 \times 10^{12} M_{\odot}$ \cite{SS98}.  Despite the smaller mass,
the proximity of the Local Group to us leads to non-negligible
contributions to the UHECR flux: about a factor $\sim$ 2 for the
$3\times 0.6$ eV model, and up to a factor of $\sim 10$ at $E \agt
10^{20}$ eV for the 1.8 eV model. The difference is primarily due to
the more efficient clustering of 1.8 eV neutrinos compared to 0.6 eV
neutrinos in the Local Group.

Our main conclusion from Fig.~6 is that the flux of UHECRs in the
Z-burst model should show significant anisotropy if $m_\nu \go 0.3$
eV, with the largest flux coming from the Virgo cluster.  For $m_\nu
\lo 0.1$ eV, on the other hand, neutrinos are too hot to cluster
appreciably even around the largest clusters in the universe, and the
UHECR flux in the Z-burst model is nearly isotropic.  We choose to
plot in Fig.~6 the flux per steradian because the angular extent of
the clusters cannot be precisely defined, but one can easily use the
approximate angular extents of the clusters listed in Table~1 to
estimate the expected anisotropy in the signal.

\section{Conclusion and Discussion}

We have introduced and tested a method based on the collisionless
Boltzmann equation to calculate the gravitational clustering of
massive neutrinos in CDM halos for realistic cosmological models.
This method is valid for currently favored models with $\Omega_{\rm
cdm} \gg \Omega_\nu$ in which the clustering of neutrinos is mostly
determined by the existing CDM halos while the clustering of the CDM
is little affected by the neutrinos.  One can then obtain the neutrino
phase space distribution by solving the collisionless Boltzmann
equation in a background potential given by the universal profile of
CDM halos from high resolution simulations.  The resulting Boltzmann
equation is linear in the neutrino density contrast and has tractable
intergral solutions that require negligible computational time in
comparison with N-body simulations.  This method has enabled us to
obtain specific predictions for the neutrino overdensity as a function
of halo radius, halo mass, and neutrino mass for a wide range of
parameters.

Our calculation shows that neutrinos with masses $\agt$ 0.3 eV can
cluster appreciably in CDM potential wells with masses $\agt 10^{13}
M_{\odot}$.  The predicted neutrino overdensity increases with both
the neutrino mass and the halo mass, ranging from $\sim 10$
for 0.3 eV neutrinos in $\sim 10^{13} M_\odot$ halos to
$\sim 1500$ for 1.8 eV neutrinos in $\sim 10^{15} M_\odot$.
Specific predictions are plotted in Figs.~2 and 3.  Neutrino
clustering has a strong impact on the Z-burst model that has been
proposed as a possible explanation for the UHECR events.  The
predicted UHECR spectrum shown in Figs.~5 and 6 depends sensitively on
the neutrino mass and overdensity, showing distinct spectral features
towards nearby galaxy clusters if $m_\nu\agt 0.3$~eV.

To illustrate the effects of neutrino mass and overdensity on the
UHECR spectrum, we have chosen to normalize the flux in Figs.~5 and
6 with the same value (i.e. $F_{\nu_{i}}(E_{\nu_{i}}^{res})= 1.7
\times 10^{-35}$ (eV m$^2$ s sr)$^{-1}$ for each flavor for the three
degenerate mass models and three times higher for the one massive
species model) instead of adjusting it by fitting individual spectrum
to existing data.  We have nonetheless included current data from the
AGASA \cite{AG} and HiRes \cite{HR} experiments in Figures~5 and 6 for
comparison.  More events are needed to discriminate the different
models and the directional dependence.  The large increase in flux
towards Virgo is an interesting signature of the Z-burst model for
upcoming experiments such as Auger \cite{AUG} and OWL \cite{OWL99}
that will provide an angular resolution of $\sim$ 1$^{\circ}$.
Experimental limits on the anisotropy would in turn imply small
neutrino inhomogeneities in the Z-burst model and can be used to place
upper bounds on the neutrino mass.

A useful constraint on the Z-burst model is provided by the Energetic
Gamma Ray Experiment Telescope (EGRET) measurement of the GeV
$\gamma$-ray background flux, which must not be exceeded by the high
energy photons produced in the Z-burst models once they cascade down
to the GeV energy range.  The result depends on the assumed redshift
evolution of the sources that produce the incident high energy
neutrinos, and on whether the sources themselves produce photons. The 
normalization of the neutrino flux cited in the previous paragraph rules out 
sources emitting a comparable flux in $\gamma$-rays because it leads to a 
conflict with the existing EGRET limits for the GeV $\gamma$-rays. For
pure neutrino sources, calculations based on particle transport codes
show that for neutrino masses of 0.1, 0.5, and 1 eV (ignoring neutrino
clustering), the EGRET bound is met for $\alpha\lo -3$, $\lo 0$, and
$\lo 3$, respectively, where the source number density evolves as
$(1+z)^\alpha$ \cite{FKR02,KKSS02}. When neutrino
clustering is taken into account, our results from Fig.~2 show that
the bound above for $m_\nu \lo 0.3$ eV should be unaffected since they
do not cluster appreciably in the Local Group.  For larger neutrinos
masses, however, we expect a less stringent bound on the source
evolution due to local neutrino clustering.  To derive quantitative
constraints would require detailed transport calculations.

The implications of the neutrino clustering results presented in this
paper extend beyond the problem of the UHECR spectrum.  For UHECR,
upcoming experimental results may converge on a spectrum that is
consistent with the GZK cutoff and would therefore not require models
such as the Z-burst.  It is also likely that the Z-burst model is not
the correct explanation for the UHECR events.  However, the
neutrino-anti-neutrino resonance scattering process remains one of few
ways to detect the relic neutrinos, as first suggested in
Ref.~\cite{W82}.  This paper has addressed neutrino clustering, a
major uncertainty in all studies concerning relic neutrinos.

\begin{acknowledgments}

We thank Ed Bertschinger, Tom Weiler, Jon Arons, Bhuvnesh Jain, and
Nick Sarbu for useful discussions.  We also thank Dmitry Semikoz, John
Beacom, and the referee for comments that have helped to improve the
manuscript.  C.-P. M. acknowledges support of an Alfred P. Sloan
Foundation Fellowship, a Cottrell Scholars Award from the Research
Corporation, and NASA grant NAG 5-12173.  A portion of the work was
carried out at the Aspen Center for Physics.  This research has made
use of the NASA/IPAC Extragalactic Database (NED) which is operated by
the Jet Propulsion Laboratory, California Institute of Technology,
under contract with the National Aeronautics and Space Administration
and the VizieR catalogue access tool, CDS, Strasbourg, France.

\end{acknowledgments}



\clearpage

\begin{figure*}
\begin{tabular}{c}
\psfig{file=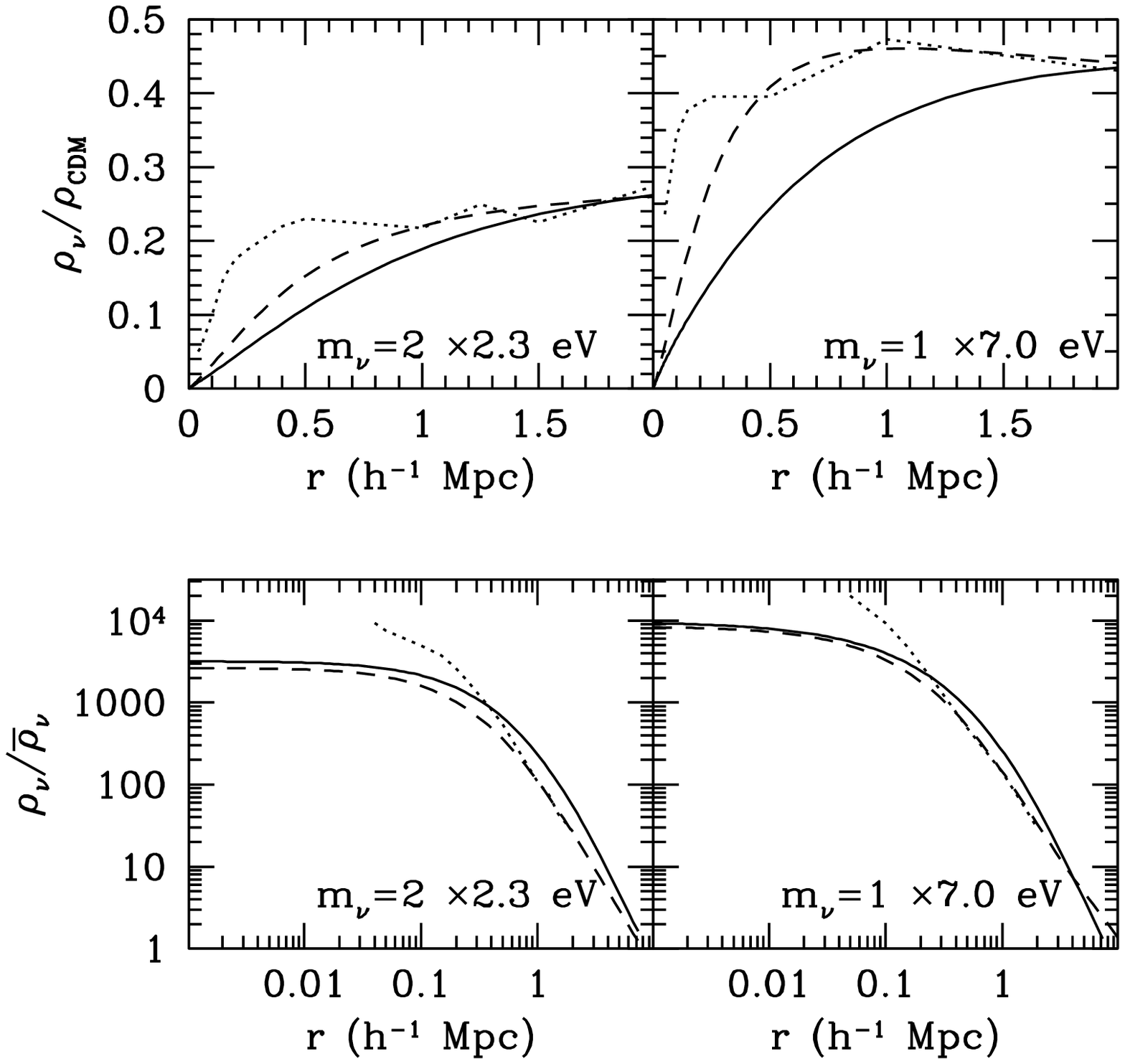,width=7.5in} 
\end{tabular}
\caption{ 
Neutrino clustering calculated with our Boltzmann approach (dashed)
vs. numerical simulations from \cite{KKPH96} (dotted).  The simulation
resolution is 62.5 h$^{-1}$ kpc.  The upper panels show the ratio of
the neutrino mass density $\rho_{\nu}$ to the CDM density $\rhocdm$ as
a function of radius for two halos of $1.3\times 10^{15} M_\odot$ in
two cosmological models.  The lower panels show
$\rho_\nu/\bar\rho_\nu$ for the same models.  The solid curves compare
neutrino clustering around CDM halos with an NFW profile \cite{NFW96}
to illustrate how neutrinos respond to different gravitational
potentials.  }
\end{figure*}

\begin{figure*}
\begin{tabular}{c}
\psfig{file=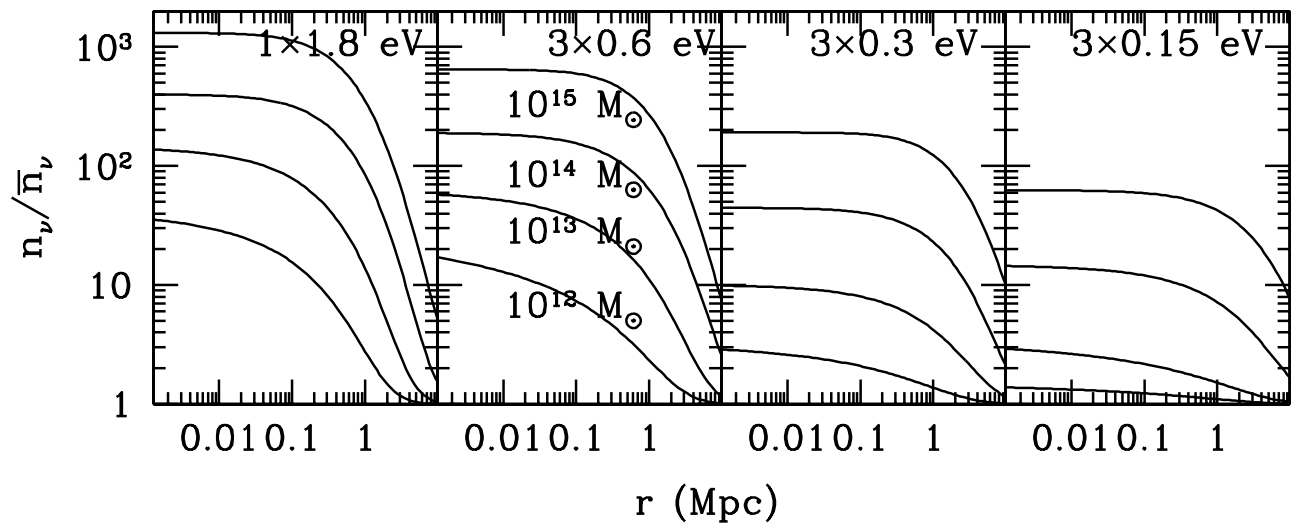,width=7.5in} 
\end{tabular}
\caption{
Total neutrino number density $n_{\nu}(r)$ as a function of halo
radius for different neutrino masses and halo masses at redshift 0.
The curves are all normalized to $\bar{n}_{\nu} \approx 112$ cm$^{-3}$
for ease of comparison.  The four panels (from left to right) show how
the clustering decreases as the neutrino mass is lowered, a result of
increasing neutrino thermal velocity and more effective free
streaming.  Within each panel, the curves show how $n_{\nu}$ decreases
as the halo mass is lowered from $10^{15}$ to $10^{12} M_\odot$, a
result of shallower gravitational wells and smaller halo velocity
dispersions compared with the neutrino thermal velocity. This figure
shows that neutrinos with $m_\nu \go 0.15$ eV cluster appreciably in
$M \go 10^{12} M_\odot$ halos.  }
\end{figure*}

\begin{figure*}
\begin{tabular}{c}
\psfig{file=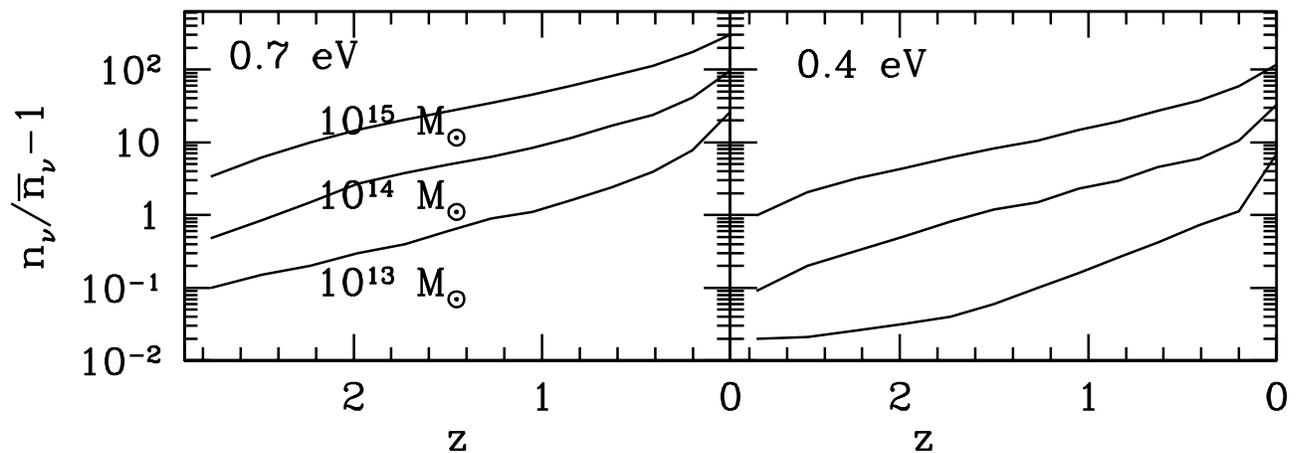,width=7.5in} 
\end{tabular}
\caption{ Time evolution of the neutrino overdensity in the inner
parts of the halos (where $n_{\nu}$ is independent of radius) for
$m_\nu=0.7$ (left) and 0.4 eV (right).  In each panel, three halo
masses $10^{15},10^{14}$, and $10^{13} M_\odot$ are shown (top down).
Neutrinos start to cluster significantly only at late times, with
$\agt$ 75 \% of the clustering taking place between $z=1$ and 0.  }
\end{figure*}

\begin{figure*}
\begin{tabular}{c}
\psfig{file=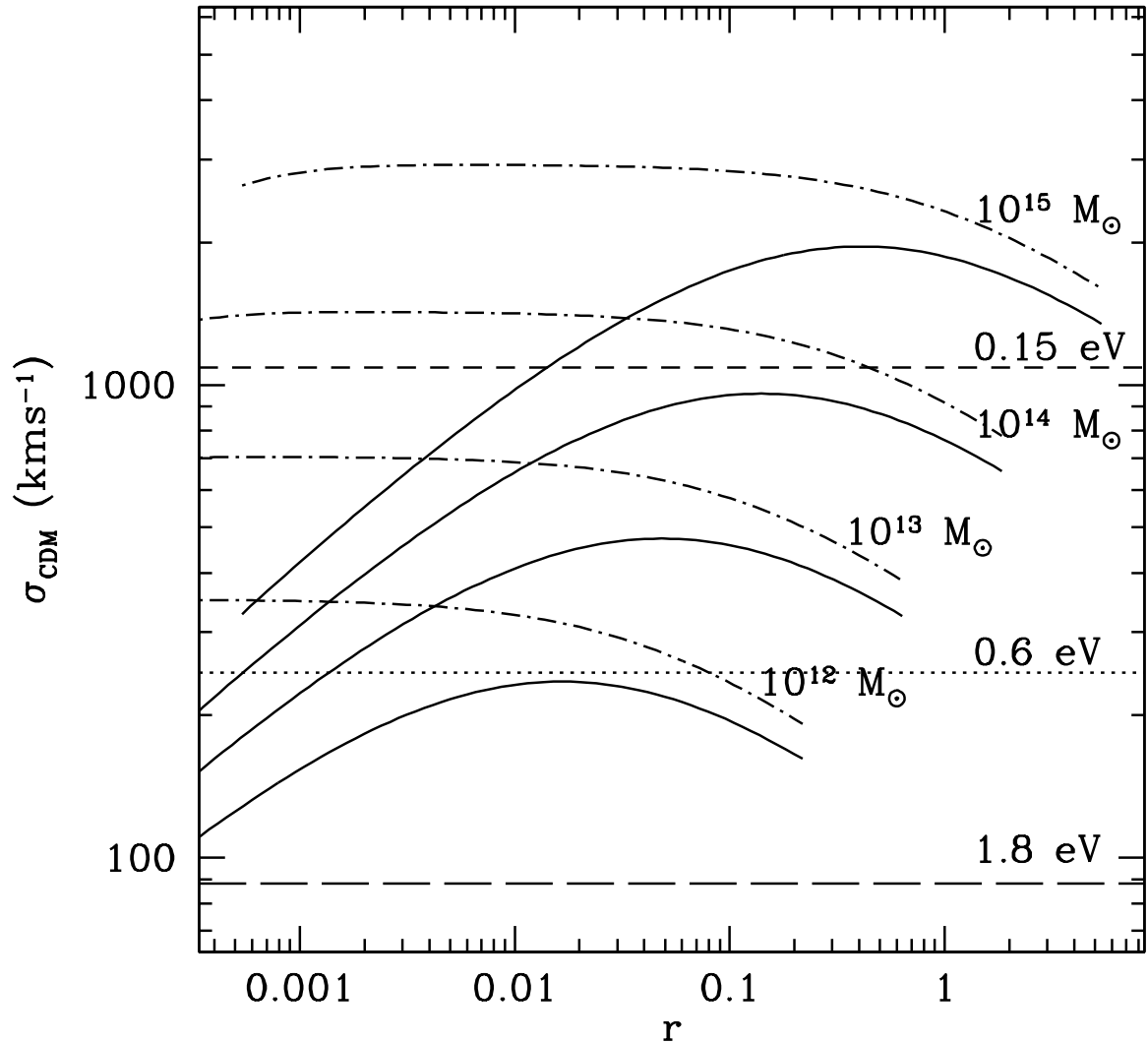,width=7.5in} 
\end{tabular}
\caption{ Velocity dispersion of NFW halos of mass $10^{15}, 10^{14}
10^{13}$, and $10^{12} M_\odot$ (top down) as a function of halo
radius.  Two velocity orbits for the halo particles are shown for
comparison: isotropic (solid) and $\beta\equiv 1-v^2_t/v^2_r=0.5$
(dot-dashed).  The horizontal lines indicate the present-day median
thermal velocity for 0.15, 0.6, and 1.8 eV neutrinos.  The values
suggest that $m_\nu \go 0.15$ eV neutrinos are cold enough to cluster
gravitationally, particularly in massive halos.}
\end{figure*}

\begin{figure*}
\begin{tabular}{c}
\psfig{file=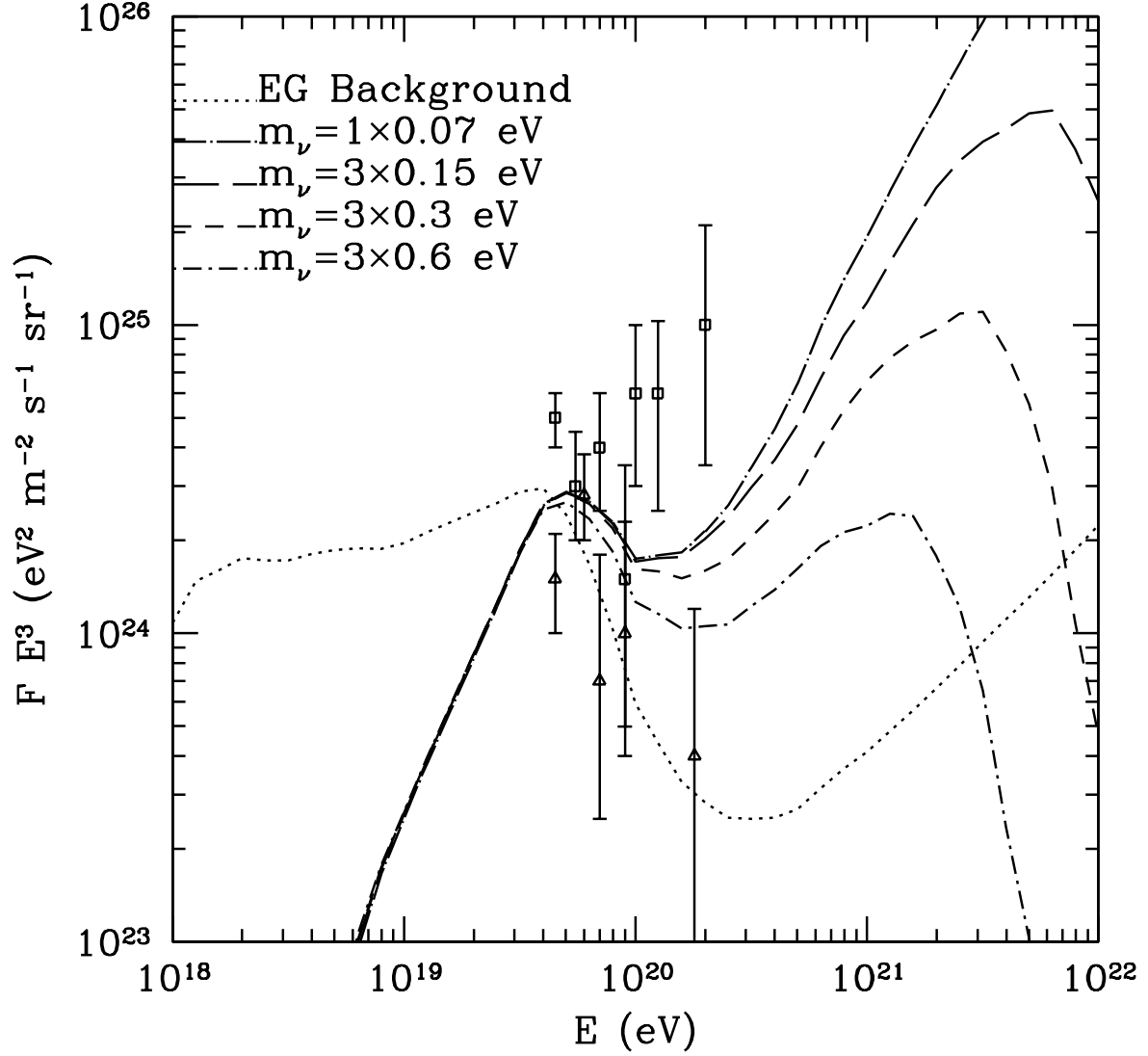,width=7.5in} 
\end{tabular}
\caption{ Predictions for the cosmic ray flux produced in the Z-burst
model when the relic neutrino density is assumed to be the big bang
uniform background density without gravitational clustering.  Three of
the four models shown assume three degenerate masses, each with 0.6,
0.3, and 0.15 eV (from bottom up); the fourth model assumes a single
massive species with 0.07 eV.  The cosmic ray spectrum of protons
originating from a uniform extragalactic background sources is shown
for comparison (dotted).  The GZK suppression in the flux is clearly
seen at $E\go 4 \times 10^{19}$ eV in all spectra.  The squares show
the current 30 UHECR events from AGASA \cite{AG}; the triangles show
the HiRes events \cite{HR}.  }
\end{figure*}

\begin{figure*}
\begin{tabular}{c}
\psfig{file=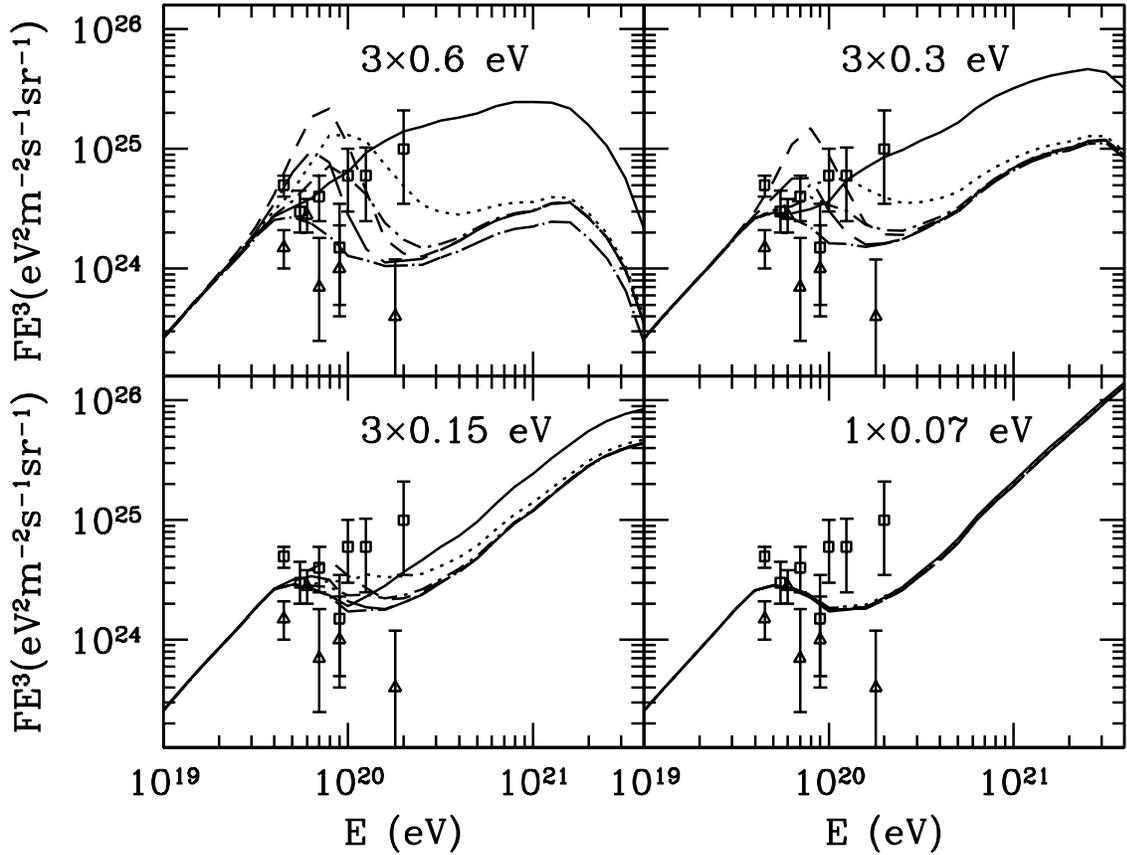,width=7.5in}
\end{tabular}
\caption{ Predictions for the cosmic ray flux produced in the Z-burst
model using realistic neutrino overdensity computed in this paper.
The four panels compare the UHECR spectrum for the same four neutrino
mass models as Fig.~5.  Within each panel, our predictions for the
spectrum towards five of the most massive clusters in the local
universe are shown: Virgo (solid), Centaurus (dotted), Hydra (dot
short-dashed), Perseus-Pisces (short dashed), and Coma (long dashed).
For comparison, the dot- long-dashed curve shows the spectrum when
neutrino clustering is ignored (i.e. the same as in Fig.~5).  For
$m_\nu \go 0.3$ eV, we predict that the Z-burst model should produce
distinct spectrum towards each line of sight.  For $m_\nu \lo 0.1$ eV,
neutrino clustering is insignificant and the spectrum is expected to
be nearly isotropic as seen in the lower right panel.  The data points
are the same as in Fig. 5.}
\end{figure*}

\end{document}